\documentclass{IOS-Book-Article} 
\usepackage{graphicx}

\begin{document}
\begin{frontmatter}

\title{Humanwashing - It Should Leave You Feeling Dirty}
\runningtitle{Humanwashing}

\author[A]{\fnms{Ben} \snm{Wilson}}
\author[B]{\fnms{Matimba} \snm{Swana}}
\author[B]{\fnms{Peter} \snm{Winter}}
and
\author[A]{\fnms{Matt} \snm{Roach}}
\address[A]{Computational Foundry, Swansea University, UK}
\address[B]{University of Bristol, UK}
\runningauthor{Wilson et al}

\begin{abstract}

The phrase `human in the loop' is increasingly used to imply a sense of safety in relation to AI decision systems. It shouldn't. There are contexts where it can be applied appropriately, but these are not in the deployed decision systems we see dominating today. Human oversight of AI decision processes is one of the most popular proposals for addressing concerns, especially about bias, discrimination, misinformation, manipulation, accountability, and transparency. But there is insufficient examination of what human oversight actually means. The question raised in this paper is whether using the metaphor of a loop does anything to assist understanding of what is required and what is achieved in a particular decision context. Indiscriminate use of the loop metaphor obscures both processes and outcomes. It enables `humanwashing', an activity analogous to `greenwashing', where writers and commentators use language primarily aimed at putting systems in the best possible light.

\end{abstract}


\end{frontmatter}

\section{Assuring decision systems}

AI decision processes are increasingly encountered in our daily lives. 
We can see around us that responses range from enthusiasm and curiosity to scepticism and aversion. 
Those who are sceptical or averse seek assurance.
Like responsible researchers, they worry especially about bias, discrimination, misinformation, manipulation, accountability, and transparency. 
One of the most popular proposals from AI developers for addressing concerns is the assurance of human oversight of AI decision processes - most commonly formulated as putting a `human in the loop'.
This is most evident in regulations and discussion about those regulations within the EU and other jurisdictions.
Having a human review machine decisions is seen as protection against system error.
But there is insufficient examination of what human oversight actually means \cite{green2022FlawsPoliciesRequiring}.
Eggert raises three problems with the popular notion of `meaningful human control' -- compliance, dignity, and responsibility \cite[p215]{eggert2024RethinkingMeaningfulHuman}.
Though these problems are identified in the context of lethal AI systems, all three are recognisably present in diverse, consequential sectors, such as healthcare AI, finance AI, vehicular AI, and aviation AI.
And yet Holzinger et al. argue that modern AI systems demand nuanced oversight mechanisms that are often underdeveloped or even lacking entirely \cite{holzinger2025HumanOversightAI}.

The background against which this is being played out is that across Europe, including the UK, authorities and businesses are keen to `unleash' the power of AI. 
They see it as a means of driving economic growth. 
The UK government's pro-innovation white paper published in 2023 defines AI in terms of adaptivity and `autonomy'. 
The latter being the feature possessed by some systems that enables them to `make decisions without the express intent or ongoing control of a human.' \cite[p24]{DSIT2023ProinnovationApproachAI} 
The white paper's miniature case studies contrast chatbot operation having `little need for human oversight or intervention' with healthcare triage in which an AI system makes recommendations to a human \cite[p26]{DSIT2023ProinnovationApproachAI}.
The implication is that human oversight or intervention should provide reassurance.
And of course the UK Information Commissioner's Office insists that `human input needs to be meaningful' in a decision-support process \cite{ico2025HowWeEnsure}.
But in many academic papers, in supplier promotion and in deployed systems it is rarely made clear how this is to be achieved.

Outside of the UK, the EU AI Act (2024) defines an AI system in similar terms – `autonomy' and adaptiveness (Article 3) \cite{eu2024Article3Definitions} - and it states the requirement of effective `human oversight' (Article 14) \cite{eu2024Article14Human}. 
But digging into what this means is complex and not yet clear in legal terms, with no sign of assurance being a foundational concept. 
Even before it was withdrawn \cite{Rankin2025EUWithdrawsProposedRules}, the AI Liability Directive was seen as problematic by citizen groups in that it protected commercial secrets rather than claimants' rights, with many applications not included - leaving the burden of proof of causality a huge obstacle \cite{madiega2023ArtificialIntelligenceLiability}. 
The exclusion from this regulation of systems in which a human played a role was an incentive to notionally describe a system as having a `human in the loop' without being any more explicit \cite{wachter2024LimitationsLoopholesEU}. 
Similarly, the EU's General Data Protection Regulation (GDPR) stipulates that right of access (Art 15) and right to an explanation (Art 22) only apply where a decision is based `solely on automated processing' \cite{eu2016GDPR,wachter2016WhyRightExplanation}.
It is likely that the vast majority of AI deployments will not fit into this category.
But what will we know about those that do have human involvement and hence don't have any GDPR protection?
It could easily be, `not much'.

The conflation of an information or feedback loop with a control or decision process has grown significantly (see section \ref{sec:language-loops} below). 
And it has allowed the language of loops to become problematic. 
Even if the loop is not intended to be taken literally, it has a powerful influence.
The language we use about AI frames perception and behaviour \cite[p349]{blythe2025artificial}
And it has implications for how society reflects on consequential outcomes if it allows side-stepping of accountability \cite{elish2019moralcrumple}.
As Green puts it, in spite of the reliance on such unspecified human oversight, the bland invocation to employ this approach rests on `an uninterrogated assumption: that people are able to effectively oversee algorithmic decision-making' \cite[p1]{green2022FlawsPoliciesRequiring}.
Mention of humans in loops, says Green, `may appear to satisfy legal and philosophical principles' but there is no empirical evidence that improvement follows \cite[p2]{green2022FlawsPoliciesRequiring}.
Worse still, `human oversight policies create a loophole that allows agencies to adopt flawed algorithms and to shirk accountability for any harms that result.' \cite[p9]{green2022FlawsPoliciesRequiring}
And these policies serve to `legitimize the use of flawed and unaccountable algorithms in government' \cite[p2]{green2022FlawsPoliciesRequiring}. 
Our ambition should be to improve our understanding of the effects of combining humans and machines in decision-making.

\section{Human oversight - ideas and realities}
\label{sec:human-oversight}

A typical definition for `human in the loop' is an AI approach that keeps `human input in the algorithmic decision-making process at some level.' \cite[p1]{valtonen2022HumanintheloopExplainableAccurate}
But this is sufficiently vague as to be practically meaningless.
Claims that putting a human `in the loop'  automatically assures oversight as well as performance improvement are so pervasive online that they dominate AI-generated search summaries of the subject.
And such summaries themselves will frequently assert that the claim arises from authoritative sources.
But according to work by Vaccaro et al \cite{vaccaro2024WhenCombinationsHumans}, human-AI combinations performed significantly worse than the best of humans or AI alone in a survey of over a hundred peer reviewed studies of decision systems \cite[p2294]{vaccaro2024WhenCombinationsHumans}. 
The authors also found that while there were gains in creation tasks, there were performance losses in decision tasks \cite[p2296]{vaccaro2024WhenCombinationsHumans}. 
In an intriguing and counter-intuitive result that provides a glimpse into the complexity of combining humans with AI, the authors found that where AI outperformed humans, there was an overall loss of performance in the combination. 
Whereas where the AI did not outperform humans, they found performance gains in the combination \cite[p2295]{vaccaro2024WhenCombinationsHumans}.
So improved models may actually lead to poorer outcomes and less oversight.

This indicates that it is insufficient to assert that synergy \cite{turchi2026SixElementsSynergy} or gains from combining humans and AI will automatically arise, merely because humans and AI bring different strengths and weaknesses to a decision process \cite{vanzoelen2023DevelopingTeamDesign}.
There is an underlying challenge here to the Hybrid Human-AI (HHAI) research community to recognise that we need a much deeper understanding of the ways in which humans and machines combine to produce decision outcomes.

The specific question raised in this paper is whether using the metaphor of a `loop' does anything to assist that deeper understanding.
If we begin with, or reinforce, a loop metaphor, does it serve us well in understanding what is required and what is achieved in a particular decision context? 
We argue that it does not.
Not only does it clarify nothing of the real and complex process of combination, but indiscriminate use of the loop metaphor obscures both processes and outcomes.
It reduces the space of combination to an apparently linear process and at the same time it enables `humanwashing', an activity analogous to `greenwashing', where writers and commentators join with commercial promoters in using language and labels aimed at putting products or systems into the best possible light without disclosing uncomfortable facts about their possible downsides.

\section{The Language of Loops}
\label{sec:language-loops}

\begin{figure}
  \centering
  \includegraphics[width=\linewidth]{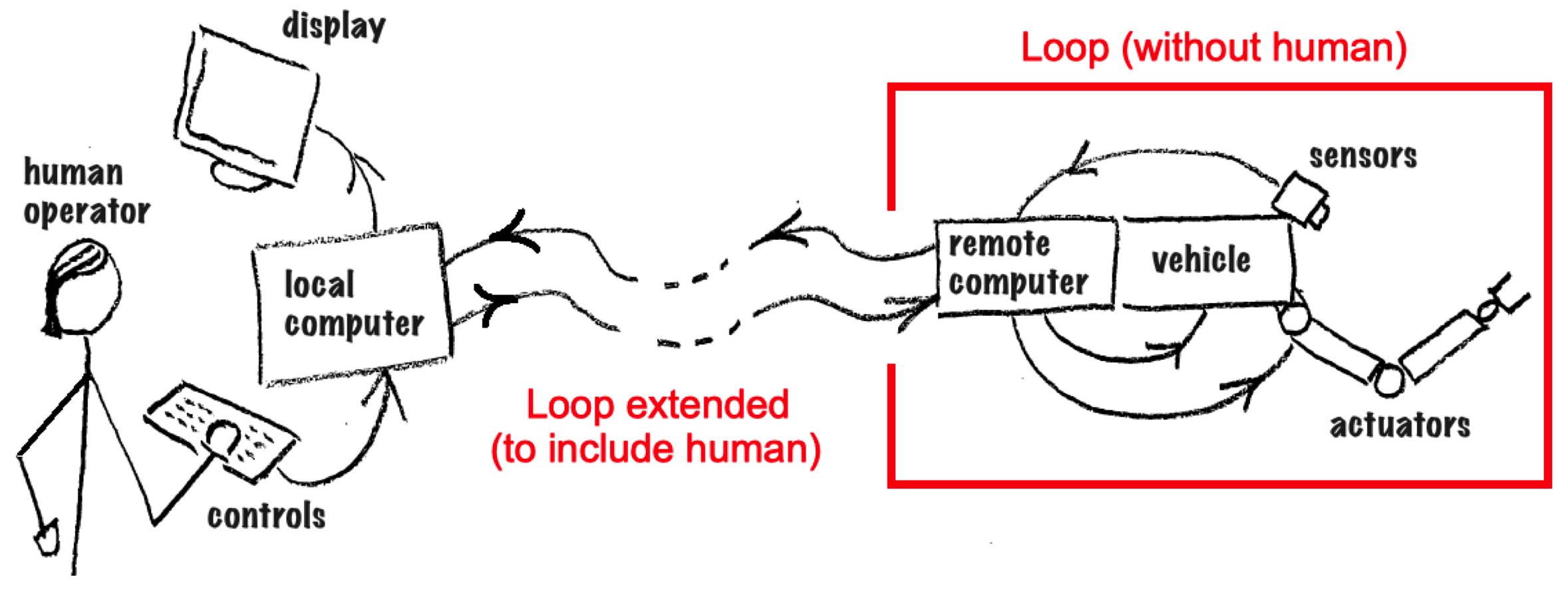}
  \caption{A real `human in the loop' system. Author facsimile of detail from Figure 1 of Sheridan et al~\cite{sheridan1978HumanComputerControl} showing how a sensor-actuator loop without a human can be extended to include a human who is thus `in the loop'. There is no analogue of such loops in decision systems.}
  \label{fig:sheridan}
\end{figure}

Sheridan et al \cite{sheridan1978HumanComputerControl} distinguish between direct teleoperation and supervisory control. 
Their definitional drawing indicating how a system capable of `decision making and control' may be monitored, operated or reprogrammed through an extended control circuit (Fig \ref{fig:sheridan}) may be one of the earliest to exhibit a loop.

It feels instructive that writers of this period distinguish monitoring and acting, `Automation of control and automation of monitoring are quite independent of one another' \cite[p13]{wiener1980FlightdeckAutomationPromises}.
Of course, control and monitoring are interlinked, and key writers emphasise that designs for complex systems, like aircraft, incorporating computerisation should particularly aim to mitigate the negative effects of automation of either on operator situation awareness  \cite{endsley1987ApplicationHumanFactors}.
But this is because each involves information flow to the machine instead of to the human - allowing the human to lose situational awareness.
Billings (1991) describes processes at different levels in order to focus on the issue of operator bandwidth, but the discussion makes clear that the differences involve moving from information to control \cite[p50]{billings1991HumancenteredAircraftAutomation}. 
There is an interesting description of `management by consent' - an approach that \textit{requires} pilot involvement and hence some level of awareness \cite[p35]{billings1991HumancenteredAircraftAutomation}.
We can distinguish, too, in the writing of Endsley and Kiris (1995) between a control loop and an information loop. 
Their discussion of poor `out-of-the-loop' performance makes clear they are primarily referring to an information loop – a means of situational awareness \cite{endsley1995OutoftheLoopPerformanceProblem}.
Parasuraman et al (2000) propose classes of functions that range from information to decisions and actions \cite[p286]{parasuraman2000ModelTypesLevels}. 

There is a significant body of research on military, aviation and vehicle systems that has employed variations on the loop metaphor.
Increasing automation and the drive to eliminate `human error' has led to systems being described as `human-on-the-loop'~\cite{vierhauser2021HazardAnalysisHumanontheloop,Taddeo2023awl}.
This is usually intended to indicate that a human \textit{could} intervene, but is not required to.
Unfortunately, placing a human in, on or even next to `the loop' in our writings does little to communicate what is actually happening in terms of combination, control or oversight.
From all of these sources, we can see an important early distinction between control and information.
The point is that automated monitoring provides information only - albeit curated information.
While automated action provides directives.
One is inherently passive, while the other is active by default.

This distinction is easily collapsed with our metaphors.
The notion of being kept `in the loop' by a friend or colleague is informational.
It carries with it the implicit expectation that action will not \textit{normally} be necessary.
And the loop metaphor works in the sense of inclusion, not the sense of a return path.
Information is within reach.
Any arc through which the information moves is seen to pass under the eyes of the included individual - often implying they are one among many individuals so included.
A control loop, by contrast, has very different implications.
The default is a decision of some kind that results in altered action.
Sheridan's sensor-controller-actuator loop is emphatically designed for more than information.
Information flows one way (from sensors) and instruction flows back the other way (to actuators).
It is a loop in the specific sense of a path that returns to its point of origin.

But this return path is precisely what is absent from the workings of decision systems.
Remote human supervision is not an inherent feature of abstract decision-making in the way it always has been in robotic system design.
Information flowing into a decision system is characterised by its diversity of origin - and the decision outcome is not routinely expected to flow back to each source.
In other words, robotic control loops - or loops of any kind - are a poor choice of metaphor for decision systems.

\section{Attend to the richness of the combination space}

Natarajan et al (2025) make a useful distinction between the contexts and functioning of two different types of decision systems. 
In their parlance, they associate them with the distinct aims of automation and collaboration respectively \cite{natarajan2025HumanintheloopAIintheloopAutomate}.
In their automation mode, AI systems drive inference while humans intervene. 
Meanwhile, in their `collaboration' mode, humans decide while AI assists with inference.
Notwithstanding the risk of anthropomorphising from the use of `collaboration' \cite{evans2023WeCollaborateWhat}, this classification appears to reflect the discrete roles and levels that appear in their different ways in Sheridan \cite[p348]{sheridan1978HumanComputerControl}, in Billings \cite[pp26-29]{billings1991HumancenteredAircraftAutomation}, in Endsley \& Kiris \cite{endsley1995OutoftheLoopPerformanceProblem} and, most richly, in Parasuraman et al \cite{parasuraman2000ModelTypesLevels}.
In each case, there is a distinction between assistance that is provided to the human by the machine and assistance that runs the other way, to the machine from the human.
What separates their `levels of automation' is usually a combination of sequencing and initiative. 
Is a machine suggestion provided first, or subsequent to a human contribution? 
Does the machine act unless halted?
Or does it require a positive instruction from the human?
These are all ideas to be considered as part of the design of the decision system.
But the combination space has many dimensions beyond default control \cite{wilson2025DimensionsHumanMachineCombination}.
Many dimensions, such as task overlap, temporal patterning and informational proximity arise from the sociotechnical context itself.
And each of these have an influence on the nature of human oversight.
Others, such as input influence and output representation coverage, are created by the system design.
They also influence oversight.
Still other dimensions are determined by a combination of the context and technical specification.
These dimensions, such as participating agents, control relations and informational overlap, are directly implicated in the way human oversight is enacted.

The design of decision systems that aim to leverage the complementary strengths of human and AI contributions must be improved by recognising the richness of the combination space.
Design must be conducted in the light of the given sociotechnical context along with factors concerning the decision task and how decision-making is framed for the human participants involved, how the AI capabilities are matched to the task and how outcomes are evaluated according to the context and the combination \cite{turchi2026SixElementsSynergy}.

\section{Can we leverage the loop metaphor for combined decision-making?}

For a distinct class of systems, we can actually identify a loop.
These are systems involving Interactive Machine Learning and are the focus of papers like those of Natarajan et al \cite{natarajan2025HumanintheloopAIintheloopAutomate}.
Some of these describe the learning process as a loop \cite{geissler2025HumanLatentLoop,stumpf2009InteractingMeaningfullyMachine}.
Others appear to carefully avoid the phrase \cite{teso2019ExplanatoryInteractiveMachine}.
In all these systems, humans provide feedback that helps the machine to improve its decision-making or classification.
This is an extension of the training phase undertaken by all machine systems.
This learning process can legitimately be considered as an iterative loop.
But we should recognise that such proposed decision systems are not yet at any state of development that brings them remotely close to deployment in consequential contexts.
And we should further recognise that learning loops are not designed to provide oversight.
Our discussion of deployed decision systems should reflect the fact that they employ inference from previously trained machines.

What are the implications for human-machine combined decision-making?
Some writers have sought to take hold of the loop metaphor and use it to re-centre human involvement in decision-making.
As against `human in the loop' they posit the idea of `computers in the loop' \cite{wiethof2021HybridIntelligenceCombining}.
Ben Shneiderman objected that `the loop’ is peripheral, so it is better to relegate the computers to `the loop' so as to centre humans \cite{shneiderman2020HumanCenteredArtificialIntelligencea}. 
Unfortunately, the re-framing attempt creates two problems.
First, it reinforces the idea that there is a meaningful loop in the inference process - something with both an outward and an inward element.
And second, it serves to flatten the diversity of combination processes and effects in precisely the place they should be explored - at the nexus of humans and machines.
We risk having our thinking about a rich, multi-dimensional space \cite{wilson2025DimensionsHumanMachineCombination} being reduced, mentally, to movement along a single-dimension - `the loop'.

\section{Can combination be achieved?}

Some writers have been emphatic about the challenges of ever successfully combining humans and machines - `machine and human decision-making are not readily compatible, making the integration of human and machine decision-making extremely complicated.' \cite[p1923]{solove2024AIALGORITHMSAWFUL}.
Others point out that user expectations may not be met because they `may or may not not reflect realities of algorithmic performance or structure' \cite[p2]{crisan2021UserExMachina}.
Or that effective human oversight is an `unrealistic expectation' \cite[p559]{vanvoorst2024ChallengesLimitationsHuman}
Yet others have delved into the variety of different interaction patterns that occur in often accidental ways. 
Creating a taxonomy of patterns is an important part of the work of understanding how and why humans and machines combine in the way they do \cite{gomez2024HumanAICollaborationNota}.
A striking feature of the results of this work is the variety of interaction patterns - none of which could readily be described as a loop.

Our argument here is that we should be much more discriminating about our use of the `loop' metaphor. Provided we attend to the richness of the combination space and holistically evaluate the consequences of combining humans and machines in decision systems, there are prospects for leveraging the complementary strengths of each. But describing this space as a `loop' doesn't help direct our thinking. Worse than this, the bland mention of a `human in the loop' is obscuring our sense of what humans do when working in combination with machines.

We therefore argue that further work is needed on the different ways that humans and machines combine in decision-making \cite{wilson2025DimensionsHumanMachineCombination}
This work should build on existing explorations of the factors that affect outcomes when humans and machines are combined, including the conditions under which such combinations improve, distort or simply displace previous human decision-making practices \cite{gajos2022PeopleEngageCognitively}.
Ultimately, this requires a careful approach to the structure, process and context of human-machine combination: who is involved, where they are positioned, what they are able to see, when they can intervene, and whether their intervention has any meaningful effect.

We intend to deepen this research by initiating closer examination of real-world implementations of human-machine combinations.
Such examination requires study of the situated work of those people who are commonly described as being `in-the-loop' as well as of the outcomes for decision subjects (for whom the phrase `in-the-loop' is rarely considered applicable). 
One strand of this work would involve quantitative research with practitioners using AI systems across different organisational and cultural contexts who are expected to intervene in, check, interpret or override decisions. 
Surveys could help identify how these actors interpret their role, what kinds of interventions they actually make including frequency of overriding AI vs acceding to it, and what language they use to describe these practices. A second strand would be qualitative and ethnographic, following the everyday work of people using AI systems in practice: observing how decisions are made, how responsibility is negotiated, and how human-machine combinations are practically accomplished in specific settings. 
This would draw on traditions in Science and Technology Studies (STS), Computer Supported Cooperative Work (CSCW) and ethnomethodology to move beyond abstract claims about oversight and towards a grounded account of what humans actually do.

Meanwhile, we urge the HHAI community to be discriminating and vigilant.
Wherever you see or hear mention of `human in the loop', ask yourself whether the context is during training or inference: is it learning or decision-making?
If it is during decision-making, ask yourself what the effect is, or what the claimed effect is, of having a human in the loop? 
Is there any work to evidence this effect? 
If it is decision-making with a human having no evident effect, it is likely to be human-washing. 
Stay away from that. 
It will distract from what is really going on. 
And it should leave you feeling dirty!

\section*{Statements and Declarations}

\subsection*{Funding}
BW acknowledges funding from the European Union. 

\noindent Funded by the European Union. Views and opinions expressed are however those of the author(s) only and do not necessarily reflect those of the European Union or the European Health and Digital Executive Agency (HaDEA). Neither the European Union nor HaDEA can be held responsible for them.
Grant Agreement no. 101120763 - TANGO. 

\subsection*{Open Access}
For the purpose of Open Access, the author has applied a CC BY licence to any Author Accepted Manuscript (AAM) version arising from this submission.

\bibliographystyle{vancouver}

\bibliography{bib_01}

\end{document}